\documentclass{caosp308}

\usepackage{graphicx}
\usepackage{tabularx}

\usepackage[]{color}

\usepackage{natbib}
\articleNo{123}
\pubyear{2020}
\volume{35}
\volnumber{3}
\firstpage{1}
\received{\today}
\accepted{\today}

\def\BibTeX{{\rm B\kern-.05em{\sc i\kern-.025em b}\kern-.08em
             T\kern-.1667em\lower.7ex\hbox{E}\kern-.125emX}}

\begin{document}

\hauthor{J.\,Merc {\it et al.}}

\title{Asteroseismology of the heartbeat star KIC\,5006817}

\author{
J.\,Merc \inst{1,2}\orcid{0000-0001-6355-2468}  
\and 
        Cs.\,Kalup \inst{3,4}
      \and
        R.\,S.\,Rathour \inst{5}\orcid{0000-0002-7448-4285}
      \and
        J.\,P.\ S\'anchez Arias \inst{6}
      \and
        P.\,G.\,Beck \inst{7,8}\orcid{0000-0003-4745-2242}
       }

\institute{
           Astronomical Institute, Faculty of Mathematics and Physics, Charles University, V Hole\v{s}ovi\v{c}k{\'a}ch 2, 180 00 Prague, Czech Republic\\ \email{jaroslav.merc@student.upjs.sk}
          \and
           Institute of Physics, Faculty of Science, P. J. \v{S}af{\'a}rik University, Park Angelinum 9, 040 01 Ko\v{s}ice, Slovak Republic
         \and
           Konkoly Observatory, Research Centre for Astronomy and Earth Sciences, Konkoly Thege 15-17, H-1121 Budapest, Hungary
         \and
           Department of Astronomy, Eötvös University, Pázmány Péter sétány 1/A, H-1171 Budapest, Hungary\\ 
         \and 
           Nicolaus Copernicus Astronomical Centre, Polish Academy of Sciences, Bartycka 18, 00-716 Warszawa, Poland \\ 
         \and
          Astronomical Institute, Academy of Sciences of Czech Republic, Fričova 298, 251 65 Ondrejov, Czech Republic \\ 
         \and
         Institut für Physik, Universität Graz, NAWI Graz, \\Universitätsplatz 5/II, 8010 Graz, Austria
         \and
           Instituto de Astrofísica de Canarias, La Laguna, Tenerife, Spain
          }

\date{March 8, 2003}

\maketitle
\vspace{-2mm}
\begin{abstract}
This paper summarizes the project work on asteroseismology at the
\hbox{ERASMUS+} GATE 2020 Summer school\footnote{\textit{'GAIA \& TESS: Tools for understanding the local Universe'}, 2020, August 7-16, Masaryk University, Brno, Czech Republic, https://gate.physics.muni.cz/} on space satellite data.
The aim was to do a global asteroseismic analysis of KIC\,5006817, and quantify its stellar properties using the high-quality, state of the art space missions data.
We employed the aperture photometry to analyze the data from the \textit{Kepler space telescope} and the \textit{Transiting Exoplanet Survey Satellite (TESS)}. Using the \textit{lightkurve} Python package, we have derived the asteroseismic parameters and calculated the stellar parameters using the scaling relations.
Our analysis of KIC\,5006817 confirmed its classification as a heartbeat binary. The rich oscillation spectrum facilitate estimating power excess ($\nu_{\rm max}$) at 145.50$\pm$0.50\,$\mu$Hz and large frequency separation ($\Delta\nu$) to be 11.63$\pm$0.10\,$\mu$Hz. Our results showed that the primary component is a low-luminosity, red-giant branch star with a mass, radius, surface gravity and luminosity of 1.53$\pm$0.07\,M$_\odot$, 5.91$\pm$0.12\,R$_\odot$, 3.08$\pm$0.01\,dex, and 19.66$\pm$0.73\,L$_\odot$, respectively. The orbital period of the system is 94.83$\pm$0.05\,d.

\end{abstract}

\section{Introduction \label{sec:introduction}}

Stellar masses and radii are the most fundamental parameters to describe a star. However, the mass and the radius of a star are difficult parameters to infer and are often found to have large uncertainties \citep[for a recent review see, e.g.][and references therein]{Serenelli2020}. Asteroseismology, the study of stellar oscillations, is a powerful technique to determine masses and radii with high accuracy as well as constrain the internal structure of a star \citep[for a recent review see, e.g.][and references therein]{Aerts2019}. A new generation of space telescopes, dedicated to obtaining photometric time series, unprecedented in photometric quality and time base, such as the NASA \textit{Kepler space telescope} \citep{Borucki2010}, has lead to the dawn of a golden age of asteroseismology. 

Such space photometric data provided the fuel for tremendous advances, particularly for the seismic investigation of stars in the advanced stellar evolution phases. Due to their extended outer convective envelopes, red-giant stars oscillate with convectively driven solar-like oscillations. Oscillation modes are classified by their restoring force.
These stochastic oscillations are mainly high order pressure-modes, leading to a very regular, comb-like pattern in the frequency domain, well explained through theory \citep{Tassoul1980}. 
The identification and quantitative description of the patterns that these modes print in the periodogram allow to determine the fundamental parameters of red-giant stars through the scaling relations \citep{1995A&A...293...87K, 2010A&A...522A...1K}. Seismically inferred masses and radii have been tested though comparison with dynamical masses of eclipsing binaries, hosting oscillating red-giant primary components \citep{Frandsen2013}. Following this approach, \cite{Gaulme2016} suggested an offset between seismic and dynamical masses of 15\%.

Furthermore, such data led to the discovery of mixed-dipole modes \citep{Beck2011, Bedding2011,Mosser2011}, being not purely pressure modes that probe the outer convective envelope, but also couple with gravity modes probing the dense central regions of the star. With this echo from the core, it was possible to constrain the density and rotational gradient between the stellar surface and the core \citep{Beck2012, Bedding2011}. In the meantime, more than 30\,000 oscillating red giants have been seismically investigated from \textit{Kepler} photometry \citep{Yu2018} and numerous sophisticated methods have been developed to exploit the complex frequency pattern of mixed modes \citep{Mosser2015, Vrard2016, Buysschaert2016}. Due to the giant's immense intrinsic luminosity and the unparalleled photometric quality of the data, asteroseismology even allows us probing stars in the galactic halo \citep{Mathur2016}.

The NASA \textit{Kepler} spacecraft data also lead to the unexpected discovery of a new class of binary systems that show tidally induced flux modulations \citep{2011ApJS..197....4W,2012ApJ...753...86T}. 
These objects, first theorized by \cite{Kumar1995}, are eccentric detached binary systems that leave a distinctive feature in their light curves, caused by large hydrostatic adjustment due to the strong gravitational distortion they experience during the periastron passage \citep[e.g.][]{Remus2012}. These stars are colloquially referred to as \textit{heartbeat} stars.

In this article, we present an asteroseismic analysis of the non-eclipsing heartbeat star KIC\,5006817. \cite{Beck2014} previously studied this system from fourteen quarters of \textit{Kepler} data covering about 1\,300 days. They found a mass of the primary component of 1.49$\pm$0.06\,M$_\odot$, and an orbital period of 94.81 days. The extensive monitoring of the radial-velocity variations with the \textsc{Hermes} spectrograph \citep{Raskin2011}, mounted to the 1.2m \textsc{Mercator} telescope on La Palma, Canary Islands, confirmed the binary nature of KIC\,5006817 and allowed them to constrain the system's eccentricity to be e=0.7.
Here, we repeat the analysis from the full and final dataset of the nominal \textit{Kepler} mission, covering more than 1\,500 days (17 quarters). We also analyze this system's latest data, obtained by the NASA \textit{TESS} mission \citep{Ricker2015}.  

\section{Data and reduction \label{sec:data}}

\subsection{\textit{Kepler} \label{sec:data:Kepler}}

With its 0.95\,m telescope, the NASA \textit{Kepler} spacecraft  \citep{Borucki2010} provided $\sim$\,4 years of continuous observation with a resolution of 4 arcsec/pixel (field of view of 105 deg$^{2}$) in both long-cadence (30 min) and short-cadence (1 min) mode. Because the spacecraft had to be rotated in order to point the solar panels towards the Sun, the time series was split into 90-day-long segments, referred to as Quarters (Q). Our target KIC\,5006817 was observed in all quarters (Q1\,-\,Q17) and the commissioning phase Q0 in the long-cadence mode. We have therefore employed all available data in the analysis. 

\begin{figure}[t!]
    \includegraphics[
    width=\textwidth]{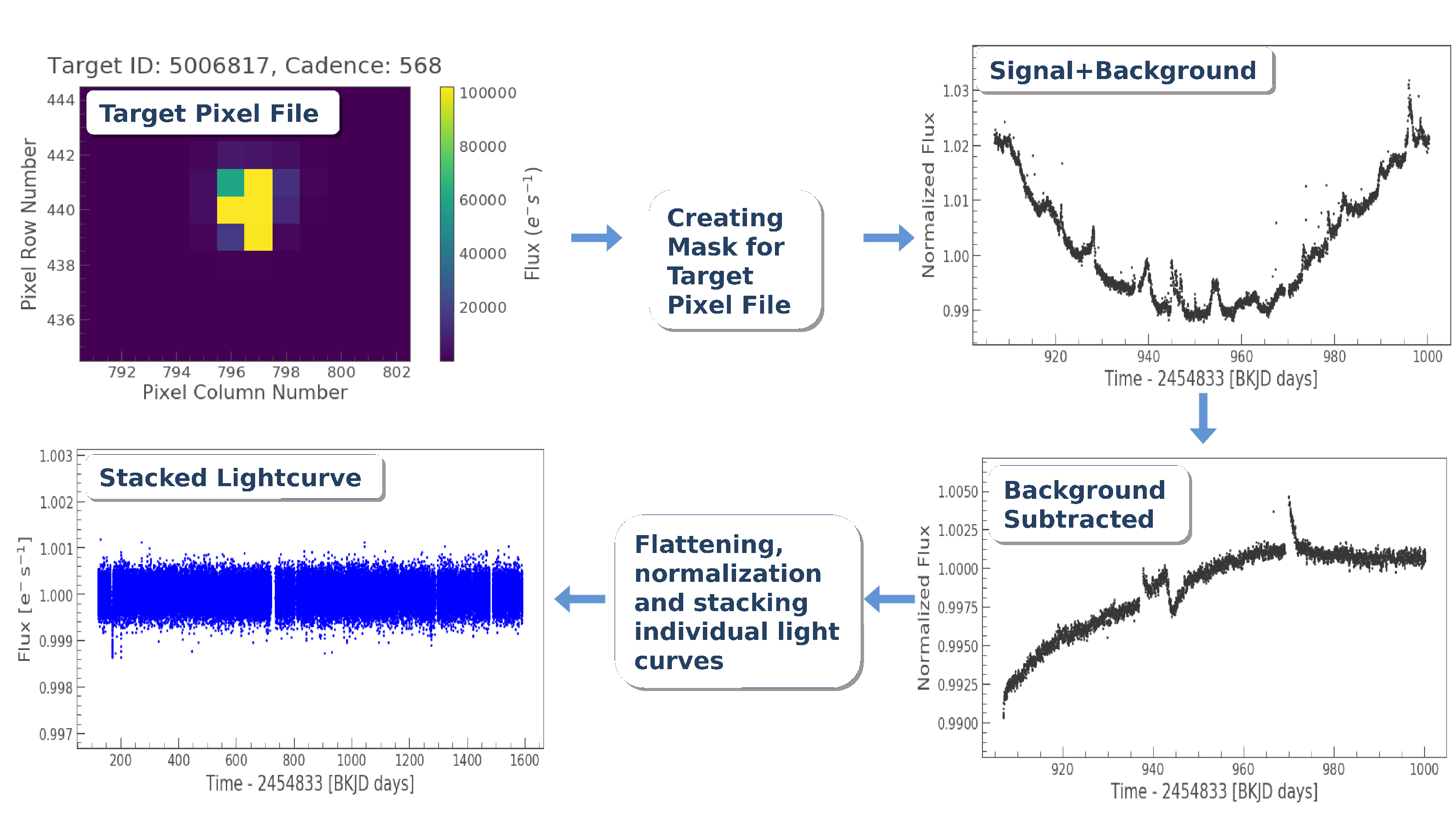}
    \caption{Analysis procedure for \textit{Kepler} data. Top left panel shows target pixel file for \textit{Kepler} Quarter 10. The top right panel shows the raw light curve with background and oscillation components. The bottom right panel shows the light curve after the background subtraction. The bottom left panel shows the full and final light curve, composed from the  normalized light curves of the individual quarters.} 
    \label{fig:procedure}
\end{figure}

We have used the \textit{lightkurve}\footnote{https://github.com/KeplerGO/lightkurve} Python package \citep{2018ascl.soft12013L,dotson2019lightkurve} to process the target pixel data of the downloaded data stamps (see Fig.\,\ref{fig:procedure}). To extract the source's flux from the target pixel files, we employed the aperture photometry. The optimal aperture was defined using a threshold feature in the package. For the present analysis, we adopted the threshold to be 25 times the median flux level to minimize the flux coming from the field stars and the background. This criterion was applied for the data extraction from all individual \textit{Kepler} quarters.

The package allowed us to remove the background signal from the light curve of the target. We have constructed a background model by selecting pixels that did not contain the target itself, any nearby bright sources, and potentially saturated pixels. Subsequently, the light curve went through a procedure of outliers removal, flattening, and normalization in each quarter individually. Finally, we stacked all the individual light curves to get the final one (Fig.\,\ref{fig:procedure}). 

There were several gaps of various length (minutes to days) in the data e.g., one missing long-cadence data point every $\sim$\,3 days due to angular momentum dump, gaps of two days due to the data download window and successive reorientation of the satellite (every $\sim$\,93 days), or other non-periodic instrumental problems, such as save-mode operation of the space craft. These regular gaps in the time domain lead to a complex alias-frequency pattern in the Fourier space \citep[see e.g.][]{Garcia2014}. To improve the spectral window, we filled the short-term gaps (few datapoints) in the data using the linear interpolation or by the Gaussian noise. Both methods gave us very similar results. This time-series data were used to obtain the periodogram to estimate asteroseismic parameters (Sect.\,\ref{sec:results}).

\subsection{{Transiting Exoplanet Survey Satellite (\textit{TESS})} \label{sec:data:TESS}}
NASA \textit{TESS} mission satellite \citep{Ricker2015} hosts four refractor telescopes with an aperture of 0.1\,m, leading to a resolution of 21 arcseconds/pixel with a combined field of view of 2304 deg$^{2}$, with two observing modes. The full field of view is co-added on board the spacecraft to full-frame images, integrated for 30 minutes. For pre-selected targets, pixel stamps around the star are stored with a 2 min short-cadence. Our target has been observed only in the long-cadence mode for one \textit{TESS} Sector 14, covering 27.5 days. We should note that the pixel scale of \textit{TESS} is significantly larger than that of \textit{Kepler}, resulting in the strong contamination of the light curve. Moreover, in the case of \textit{TESS}, the source flux is significantly affected by the stray light from the Moon and the Earth.

The reduction procedure applied to \textit{TESS} data was similar to \textit{Kepler} data, discussed in the previous section. However, in addition to aperture photometry, we have employed an alternative approach to cope with the complicated background variations and contamination, using the algorithms implemented by Andr\'as P\'al with various tasks of the FITSH package \citep{pal2012fitsh}. As we will discuss in Sect.\,\ref{sec:results:TESS}, the target is too faint for a significant detection of the asteroseismic signal with \textit{TESS}.

    \section{Binary analysis}
\label{sec:binary}

\begin{figure}[t!]
   \centering
    \includegraphics[width=0.85\textwidth]{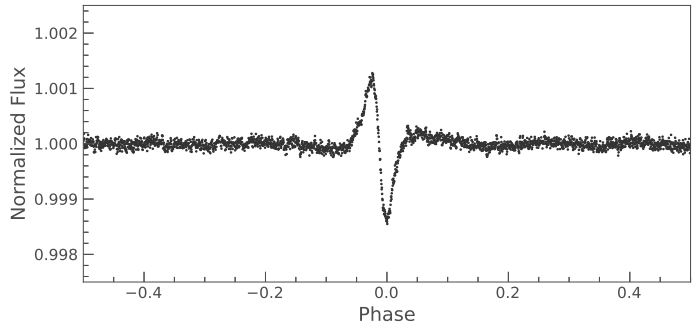}
    \caption{\textit{Kepler} light curve of KIC\,5006817 phased with the period of 94.83\,d. The heartbeat feature caused by the gravitational distortion of the primary during the periastron passage is well visible.}
    \label{fig:heartbeat}
\end{figure}

KIC\,5006817 is an eccentric binary system. Therefore, we have aimed to obtain the orbital period of the system using available photometric data. We have used the \textit{Kepler} light curve, as the \textit{TESS} data covers only short time interval of 27.5 days. For this task, the extracted \textit{Kepler} light curve was flattened using an in-built function of the \textit{lightkurve} package employing the Savitzky-Golay filter \citep{savitzky1964smoothing}, in order to remove the long-term trends in the data. The size of the flattening window was set to $\sim$\,17 days. Using this approach, we were able to preserve the heartbeat feature in the light curve, which was important for the following period analysis. We have used the Lomb-Scargle periodogram to obtain the periodogram and identify the system's orbital period to be 94.83$\pm$0.05\,d. The phase-folded light curve with this period is shown in Fig.\,\ref{fig:heartbeat}, the heartbeat feature is well visible. Our value of the orbital period is in good agreement with the work of \citet{Beck2014}, who obtained the value of 94.812$\pm$0.002\,d.

\section{Seismic analysis \label{sec:results}}

\subsection{\textit{Kepler} \label{sec:results:Kepler}}

\begin{figure}[t!]
 \centering   
    \includegraphics[width=0.85\textwidth]{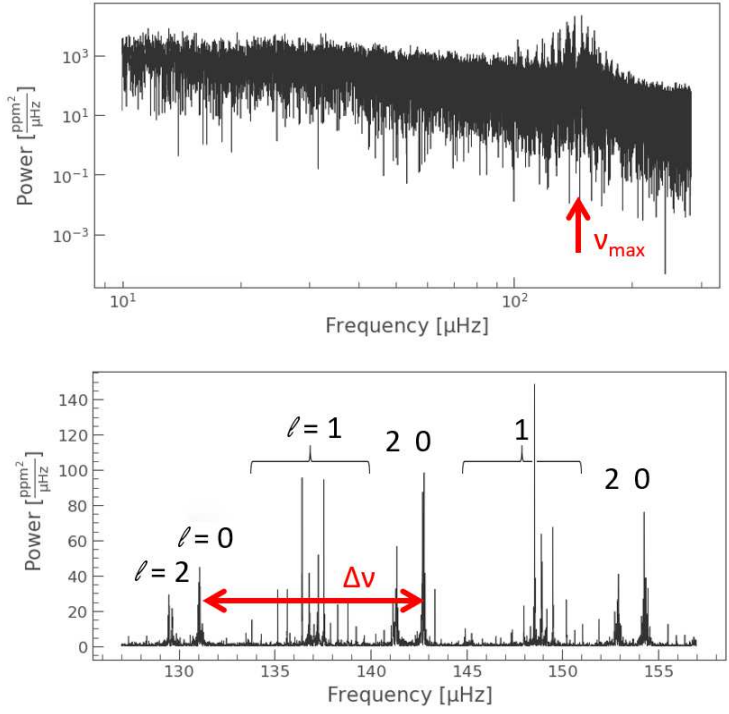}
    \caption{Asteroseismic parameters of KIC\,5006817. The upper panel shows the Power Spectral Density of the \textit{Kepler} light curve of
KIC\,5006817. The position of the peak frequency of the power excess $\nu_{\rm max}$ is shown by the red arrow. The lower panel shows a zoom into the power excess. A large frequency separation $\Delta\nu$ is shown by the red arrow. The spherical degree of the excited modes is indicated by $\ell$.
}
 \label{fig:periodogram}
\end{figure}

For the frequency analysis, we have used the light curves flattened with the size of the flattening window decreased to $\sim$\,2 days in order to remove the heartbeat feature. This filter width is small enough to remove the tidally induced flux modulation, but preserves the oscillation signal whose average period is on the order of 1.5\,hours. The individual light curves from all available quarters were then stacked together to produce the final dataset (see Sect. \ref{sec:data}). The power spectral density (shown in Fig.\,\ref{fig:periodogram}) was calculated using the Lomb-Scargle method using a built-in function in the \textit{lightkurve} package. To obtain the peak frequency of the excess of oscillation power $\nu_{\rm max}$, we have fitted the resulting periodogram by Gaussian envelope together with the power-law component to account for rotation and granulation and constant offset for the photon-noise background. To obtain the large frequency separation $\Delta\nu$, we divided the power spectral density by multi-component background model to normalize it and used the autocorrelation method \citep[see e.g.][]{2009A&A...508..877M}. Our resulting values are $\nu_{\rm max}$\,=\,145.50$\pm$0.50\,$\mu$Hz and $\Delta\nu$\,=\,11.63$\pm$0.10\,$\mu$Hz. We assume similar uncertainties of the seismic parameters as \citet[][]{2018A&A...616A.104K}, as no uncertainties have been estimated using the methods presented above.

The asteroseismic parameters obtained above are connected with the astrophysical parameters of the star. The power excess $\nu_{\rm max}$ is proportional to the surface acceleration ($\log{g}$), while the large frequency separation $\Delta\nu$ is corresponding to the speed of sound and hence the mean density of the star. The mass and radius of the star are calculated using the following scaling relations \citep{1995A&A...293...87K, 2010A&A...522A...1K}:
\begin{equation}
\frac{R_*}{R_\odot} = \frac{\nu_{\rm max}}{\nu_{\rm max}^\odot}\times\left(\frac{\Delta\nu}{\Delta\nu_\odot}\right)^{-2}\times\sqrt{\frac{T_{\rm eff}}{T_{\rm eff}^\odot}}
\end{equation}

\begin{equation}
\frac{M_*}{M_\odot} = \left(\frac{R_*}{R_\odot}\right)^{3}\times\left(\frac{\Delta\nu}{\Delta\nu_\odot}\right)^{2}
\end{equation}

\begin{equation}
\frac{L_*}{L_\odot} = \left(\frac{\nu_{\rm max}}{\nu_{\rm max}^\odot}\right)^2\times\left(\frac{\Delta\nu}{\Delta\nu_\odot}\right)^{-4}\times\left(\frac{T_{\rm eff}}{T_{\rm eff}^\odot}\right)^5
\end{equation}
where $R_*$, $M_*$, and $L_*$ are radius, mass, and luminosity of the studied star, respectively. For the calculation, we have adopted the $T_{\rm eff}$ = 5000K obtained using the high-resolution spectroscopy by \citet{Beck2014}. The solar values are from \citet{2011ApJ...743..143H} and \citet{2016AJ....152...41P}. The obtained parameters are as follows: $\log{g}$\,=\,3.08$\pm$0.01, R$_*$\,=\,5.91$\pm$0.12\,R$_\odot$, M$_*$\,=\,1.53$\pm$0.07\,M$_\odot$, L$_*$\,=\,19.66$\pm$0.73\,L$_\odot$.

\subsection{\textit{TESS} \label{sec:results:TESS}}
We have intended to repeat the same analysis as described in the previous section, also for data from the \textit{TESS} satellite. The target was observed in a single sector with a 30 min cadence. We have extracted the light curve from the full-frame images using two independent methods, the aperture photometry and the differential image analysis (see Sect.\,\ref{sec:data:TESS}). The signal to noise ratio of both resulting periodograms did not allow to reliably detect the oscillations and, therefore, to obtain the astrophysical parameters of the star. The target seems to be too faint \citep[V-band magnitude 11.15$\pm$0.08;][]{hog2000tycho} for this kind of analysis using the data obtained in 30 min cadence in a single sector. The first results on asteroseismology of the red giants using single \textit{TESS} sectors by \citet{2020ApJ...889L..34A} reported high signal-to-noise detections of the power excess of giants between 6$^{\rm th}$ and 8$^{\rm th}$ visual magnitude.
For stars in the continuous viewing zone \citep{Mackereth2020}, this limit is lower due to the increased time base of one year.

\section{Age of KIC\,5006817}
\begin{figure}[t!]
\vspace{-5mm}
\centering
    \includegraphics[width=0.85\textwidth]{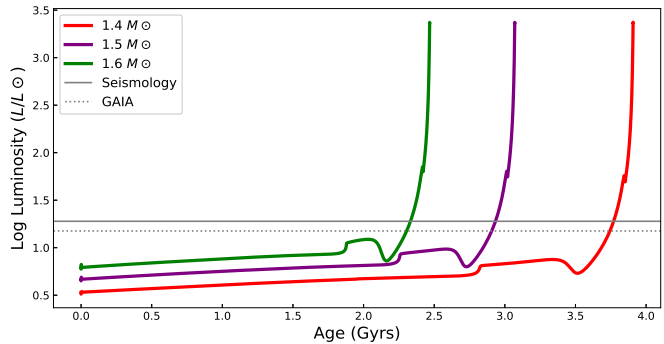}
    \caption{Luminosity-age diagram showing MESA evolutionary tracks of 1.4 (red), 1.5 (purple) and 1.6\,$M_{\odot}$ (green) with solar composition.
    The asteroseismology and \textit{Gaia} estimated luminosity marked by solid and dotted lines, respectively, are the adopted limits to estimate the age of KIC\,5006817.}
    \label{fig:mesa}
\end{figure}

Based on the seismically determined mass, we calculated a grid of simple stellar evolutionary tracks, using the MESA code \citep[\textit{Modules for Experiments in Stellar Astrophysics},][and references therein]{paxton2019modules}. The grid was computed for solar metallicity and stellar masses between 1.4, 1.5 and 1.6\,$M_{\odot}$, which roughly corresponds to the one-sigma range of the mass uncertainty. The initial metal fraction composition and opacity profiles are used as per \cite{asplund2009chemical} prescription. 

As it can be seen from Fig.\,\ref{fig:mesa}, the red-giant primary of the system belongs to the low-luminosity red-giant stars and falls between two interesting evolutionary stages. KIC\,5006817 has already passed the first dredge-up phase, when the convective envelope reaches its maximum penetration into the stellar structure. It still has to pass the phase of the luminosity bump, when the hydrogen-burning shell hits the metallicity discontinuity, left behind from the convective envelope when it started to move outwards after the first dredge-up.

As a rough estimate for the age of the red giant, we report the range of the instrumental age of MESA models, when the evolutionary track for a given mass equals the luminosity of 19.66 and 14.971 solar luminosities, found from asteroseismology (Sect.\,\ref{sec:results:Kepler}) as well as from \textit{Gaia} measurements \citep{Gaia2018}, respectively (see Fig.\,\ref{fig:mesa}). The difference between these two loci on the red-giant branch is very small as the red-giant branch is a phase of fast evolution. 
This indicates that the age of the system is somewhere between $\sim$2.3 and $\sim$3.7 Gyrs, assuming a 1.6 and 1.4 solar-mass star. We note that this is just a rough estimate, as the age of the model depends on details of the model and the included physics. The relatively large range originates from the different periods, the star spends on the main sequence. The duration of quiescent hydrogen-core burning is highly mass dependent, as nicely illustrated in Fig.\,\ref{fig:mesa}

\begin{table}[t!]
\centering
\caption{Comparison of our asteroseismic values with the literature.}
\begin{tabularx}{\textwidth}{c|>{\centering\arraybackslash}X>{\centering\arraybackslash}X>{\centering\arraybackslash}X>{\centering\arraybackslash}X}
 \hline\hline
 & This work & \citeauthor{Beck2014} & \citeauthor{Yu2018} & \citeauthor{Gaulme2020} \\
 &  & (\citeyear{Beck2014}) & (\citeyear{Yu2018}) & (\citeyear{Gaulme2020}) \\
\hline
$\nu_{\rm max}$ [$\mu$Hz] & 145.50$\pm$0.50 & 145.9$\pm$0.5 & 145.48$\pm$0.83 & 146.44$\pm$0.26 \\
$\Delta\nu$ [$\mu$Hz] & 11.63$\pm$0.10 & 11.64$\pm$0.01 & 11.620$\pm$0.017 & 11.68$\pm$0.05 \\
M$_*$ [M$_\odot$] & 1.53$\pm$0.07 & 1.49$\pm$0.06 & 1.47$\pm$0.07 & 1.39$\pm$0.05\\
R$_*$ [R$_\odot$] & 5.91$\pm$0.12 & 5.84$\pm$0.009 & 5.83$\pm$0.10 & 5.67$\pm$0.08\\
L$_*$ [L$_\odot$] & 19.66$\pm$0.73 & 19$\pm$3 & 16.87$\pm$0.61 & 15.96$\pm$0.51 \\
$\log{g}$ [dex] & 3.08$\pm$0.01 & 3.08 & 3.07$\pm$0.01 & 3.07$\pm$0.01\\
\hline
\end{tabularx}
\label{tab:asteroseismic}
\vspace{7mm}
\centering
\caption{Comparison of our astrophysical values with the non-seimic values in the literature, the Cannon \citep{2016ApJ...823..114N}, the \textit{Gaia} DR2 \citep{Gaia2018}, and the StarHorse catalog \citep{2019A&A...628A..94A}.}
\begin{tabularx}{\textwidth}{c|>{\centering\arraybackslash}X>{\centering\arraybackslash}X>{\centering\arraybackslash}X>{\centering\arraybackslash}X}
 \hline\hline
 & This work & Cannon & \textit{Gaia} DR2 & StarHorse \\
\hline
M$_*$ [M$_\odot$]  & 1.53$\pm$0.07 & 1.59$\pm 0.015$ & - & 1.15$_{-0.14}^{+0.19}$\\
R$_*$ [R$_\odot$] & 5.91$\pm$0.12 & - & 5.46$_{-0.133}^{+0.116}$ & - \\
L$_*$ [L$_\odot$] & 19.66$\pm$0.73 & - & 14.971$\pm 0.563$ & -\\
$\log{g}$ [dex] & 3.08$\pm$0.01 & 2.96$\pm 0.007$ & - & 3.00$_{-0.05}^{+0.07}$\\
\hline
\end{tabularx}
\label{tab:nonseismic}
\end{table}

\section{Discussion \label{sec:discussion}}

KIC\,5006817 has been studied by several authors in the recent decade. In Table\,\ref{tab:asteroseismic}, we have compared our asteroseismic values and the calculated astrophysical parameters of the target with the values from the literature. The first comprehensive study of the star was performed by \cite{Beck2014} based on \textit{Kepler} Q0\,-\,Q14 measurements. We also mention an analysis focused on the characterization of the solar-like oscillations and granulation in 16\,094 oscillating red giants of \citet{Yu2018}, and a recent paper by \cite{Gaulme2020}, in which the authors analyzed the asteroseismic and rotational parameters of about 4\,500 relatively bright red giants observed by \textit{Kepler}. As shown in Table\,\ref{tab:asteroseismic}, our results show good agreement with the literature within the uncertainties. 

We have also shown (in Table\,\ref{tab:nonseismic}) the comparison of the astrophysical parameters of KIC\,50068187 (mass, radius, luminosity, and $\log{g}$) obtained in this work, with the non-seismic values from the literature. The data from \textit{Gaia} DR2 \citep{Gaia2018} are based on the \textit{Gaia} broadband photometry and the measured parallax, while values in the StarHorse catalog \citep{Starhorse2020} are based, in addition to \textit{Gaia} measurements, also on the photometry from 2MASS and \textit{WISE}. \textit{The Cannon} catalog \citep{2016ApJ...823..114N} is based on the probabilistic model of stellar spectra and the spectra from the APOGEE survey. These independent approaches are also consistent with our values from asteroseismology.

\section{Conclusions \label{sec:conclusions}}
In this work, we performed an asteroseismic analysis of the \textit{Kepler} and \textit{TESS} light curves of the heartbeat star KIC\,5006817. Reasonable asteroseismic parameters, $\nu_{\rm max}$ of the power excess, and the large frequency separation $\Delta\nu$, were obtained only from \textit{Kepler} data, as the target seemed to be too faint for significant detection of oscillations using a single \textit{TESS} sector data obtained in 30 min cadence. Using the scaling relations, we calculated the mass of 1.5 M$_\odot$, radius 5.9 R$_\odot$, surface gravity 3.08 dex, and luminosity 19.7 L$_\odot$ for the primary. Analysis of the MESA evolutionary paths calculated for various masses confirmed that the red giant in KIC\,5006817 is an RGB star. The heartbeat feature detected in the \textit{Kepler} light curve allowed us to obtain the orbital period of 94.8 days.

\acknowledgements
We are thankful to an anonymous referee for the comments and suggestions improving the manuscript. The authors thank Dr. Marek Skarka, Dr. Petr Kabath, and the organizing committee for organizing this summer school in the difficult summer of 2020. We thank the people behind the \textit{Kepler}, \textit{Tess}, and \textit{Gaia} space missions, whose data we were using in this paper. 

The authors acknowledge the support from ERASMUS+ grant number 2017-1-CZ01-KA203-035562. J.M. was supported by the \textit{Charles University}, project GA UK No. 890120 and by the internal grant VVGS-PF-2019-1047 of the \textit{Faculty of Science, P. J. \v{S}af\'{a}rik University in Ko\v{s}ice}, Cs.K. acknowledges the support provided from the \'UNKP-20-2 New  National  Excellence Program of the Ministry of Human Capacities, and the LP2018-7/2020 grant of the Hungarian Academy of Sciences. R.S.R. is supported by NCN-funded Sonata Bis grant under research project no: 2018/30/E/ST9/00598. This work was supported by NAWI Graz.

This paper includes data collected by the \textit{Kepler} and the \textit{TESS} missions. Funding for the \textit{Kepler} mission is provided by the NASA Science Mission directorate, for the \textit{TESS} mission is provided by the NASA Explorer Program. This work also has made use of data from the European Space Agency (ESA) mission \textit{Gaia}, processed by the \textit{Gaia} Data Processing and Analysis Consortium (DPAC). Funding for the DPAC has been provided by national institutions, in particular, the institutions participating in the \textit{Gaia} Multilateral Agreement.

\bibliography{bibliography}

\end{document}